INDIAN STATISTICAL INSTITUTE

# COMPONENT BASED DEVELOPMENT

## APPLICATION IN SOFTWARE ENGINEERING

DEBAYAN BOSE

Component Based Development

1. **Introduction:**

   Concept of re-use is not a rare phenomenon in core engineering branches. But the same concept in software engineering context has been introduced in early days of computing but its approach was ad hoc. But the introduction of object oriented programming with some advancement explores mew areas of software engineering. Today complex, high quality software systems are built efficiently using component based approach in a short period of time. But a number of questions arise about the feasibility of the component approach. But the concept of CBD successfully answers the arising questions. The importance of Component Based development lies in its efficiency. It takes only a few minutes to assemble the stereo system because the components are designed to be integrated with ease. Although software is considerably more complex, it follows that component-based systems are easier to assemble and therefore less costly to build than systems constructed from discrete parts. In addition, CBSE encourages the use of predictable architectural patterns and standard software infrastructure, thereby leading to a higher-quality result.

2. **History of Component Based Development**

   The idea that software should be componentized - built from prefabricated components - first became prominent with Douglas McIlroy's address at the NATO conference on software engineering in Garmisch, Germany, 1968, titled Mass Produced Software Components. The conference set out to counter the so-called software crisis. McIlroy's subsequent inclusion of pipes and filters into the Unix operating system was the first implementation of an infrastructure for this idea. Brad Cox of Stepstone largely defined the modern concept of a software component. He called them Software ICs and set out to create an infrastructure and market for these components by inventing the Objective-C programming language. IBM led the path with their System Object Model (SOM) in the early 1990s. Some claim that Microsoft paved the way for actual deployment of component software with OLE and COM. As of 2010 many successful software component models exist.

3. **Definition of Component Based Software Engineering:**

   *Component-based software engineering* (CBSE) is a process that emphasizes the design and construction of computer-based systems using reusable software "components." Clements [CLE95] describes CBSE in the following way: CBSE is changing the way large software systems are developed. CBSE embodies the "buy, don't build" philosophy espoused by Fred Brooks and others. In the same way that early subroutines liberated the programmer from thinking about details, CBSE shifts the emphasis from programming software to composing software systems. Implementation has given way to integration as the focus. At its foundation is the assumption that there is sufficient commonality in many large software systems to justify developing reusable components to exploit and satisfy that commonality. Component-based software development approach is based on the idea to develop software systems by selecting appropriate off-the-shelf components and then to assemble them with a well-defined software architecture.

**Component-based software engineering (CBSE)** is a branch of software engineering which emphasizes the separation of concerns in respect of the wide-ranging functionality available throughout a given software system. This practice aims to bring about an equally wide-ranging degree of benefits in both the short-term and the long-term for the software itself and for organizations that sponsor such software.

Software engineers regard components as part of the starting platform for service-orientation. Components play this role, for example, in Web Services, and more recently, in Service-Oriented Architecture (SOA) - whereby a component is converted into a *service* and subsequently inherits further characteristics beyond that of an ordinary component.

Components can produce events or consume events and can be used for event driven architecture (EDA).

4. **Objectives of Component Based Software Engineering:**

   The main objectives of component based software engineering are given below.

   a) Reduction of cost and time for building large and complicated systems: main objective of Component based approach is to build complicated software systems using off the shelf component so that the time to build the software diminish drastically. The cost effectiveness of the current method can be analysed using function point or other methods.
   b) Improving the quality of the software: The quality of the software can be improved by improving the quality of the component. Though the concept is not true in general. Sometimes quality of the assembled systems may not be directly related to quality of the component in sense that improving the quality of the component does not necessarily imply the improvement of the systems.

c) Detection of defect within the systems: Component approach helps the system to find its defect readily by testing the components. But the source of defects is difficult to find in case of component development approach.

5. **Examples of Component Based Approach:**

   To start with, let us give an example of simple stereo systems which consists of woofer, sub-woofer, sound box etc. if someone wants to build the stereo systems by assembling the off the shelf components like sound box etc then he might be experiencing some advantage than those who build the system from the basic circuit. In-fact, now a day, all the simple and complicated systems are built using the component approach where some components are already developed by some developer and they are stored in the library for their re-use. The main concept is that those off the shelf components need not be changed for their respective purposes. If we want modify some of the components to fit according to their applicability, we need to rebuild it and store them in the library.

   Example of some complex system, where each components itself can be viewed as a system, is Naval Combat system. The system consists of some radar for detection, Helicopter, submarine, Rocket Launcher, some fighter plane etc. Each component here is some big systems and association is also very complicated in such cases.

6. **Software Components:**

   In most engineering disciplines, systems are designed by composing existing components that have been used in other systems. Software engineering has been more focused on original development but it is now recognised that to achieve better software, more quickly and at lower cost, we need to adopt a design process that is based on *systematic "reuse".*

   So, typically, components are rather small but independent parts of a system. But a large system as a whole can be seen as a component as well. It is important to recognize components are runtime entities. They exist while the system is running, in fact: the system consists of components, and is a component itself. Components are not just design entities like classes in object-orientation are. As said earlier, at his moment everybody is raving about components, and seems to expect a lot from it. What is expected form components, and why is everybody that enthusiast? The expected advantages to be derived from the application of components are summarized in later sections of this report.
   Components provide a service without regard to where the component is executing or its programming language. A component is an independent executable entity that can be made up of one or more executable objects. The component interface is published and all interactions are through the published interface. Components can range in size from simple functions to entire application systems

7. **Engineering of Component Based Systems**:
   On the surface, CBSE seems quite similar to conventional or object-oriented software engineering. The process begins when a software team establishes requirements for the system to be built using conventional requirements elicitation techniques. An architectural design is established, but rather than moving immediately into more detailed design tasks, the team examines requirements to determine what subset is directly amenable to *composition,* rather than construction. That is, the team asks the following questions for each system requirement:
   - Are commercial off the shelf components available to implement the requirement?
   - Are internally reusable components available to implement the requirement?
   - Are the interfaces of the available components compatible with the architecture of the system to be built?

The team attempts to modify or remove those system requirements that cannot be implemented with COTS or in-house components.1 If the requirement(s) cannot be changed or deleted, conventional or object-oriented software engineering methods are applied to develop those new components that must be engineered to meet the requirement(s). But for those requirements that are addressed with available components, a different set of software engineering activities commences: Component Qualification, Component Adaptation, Component Composition and Component Update. Now let us focus on few basic definition regarding software components.

*Component*—It is a nontrivial, nearly independent, and replaceable part of a system that fulfills a clear function in the context of a well-defined architecture.

*Run-time software component*—It is a dynamic bindable package of one or more programs managed as a unit and accessed through documented interfaces that can be discovered in run time.

*Software component*—It is a unit of composition with contractually specified and explicit context dependencies only.

• *Business component*—It is the software implementation of an "autonomous" business concept or business process.

In addition to these descriptions, software components can also be characterized based on their use in the CBSE process. In addition to COTS components, the CBSE process yields:

*Qualified components*—assessed by software engineers to ensure that not only functionality, but performance, reliability, usability, and other quality factors (Chapter 19) conform to the requirements of the system or product to be built.

• *Adapted components*—adapted to modify (also called *mask* or *wrap*) [BRO96] unwanted or undesirable characteristics.

• *Assembled components*—integrated into an architectural style and interconnected with an appropriate infrastructure that allows the components to be coordinated and managed effectively.

• *Updated components*—replacing existing software as new versions of components become available.

8. **Component Based Software Engineering Processes:**

A "component-based development model" (Figure 1) was used to illustrate how a library of reusable "candidate components" can be integrated into a typical evolutionary process model. The CBSE process, however, must be characterized in a manner that not only identifies candidate components but also qualifies each component's interface, adapts components to remove architectural mismatches, assembles components into a selected architectural style, and updates components as requirements for the system change.

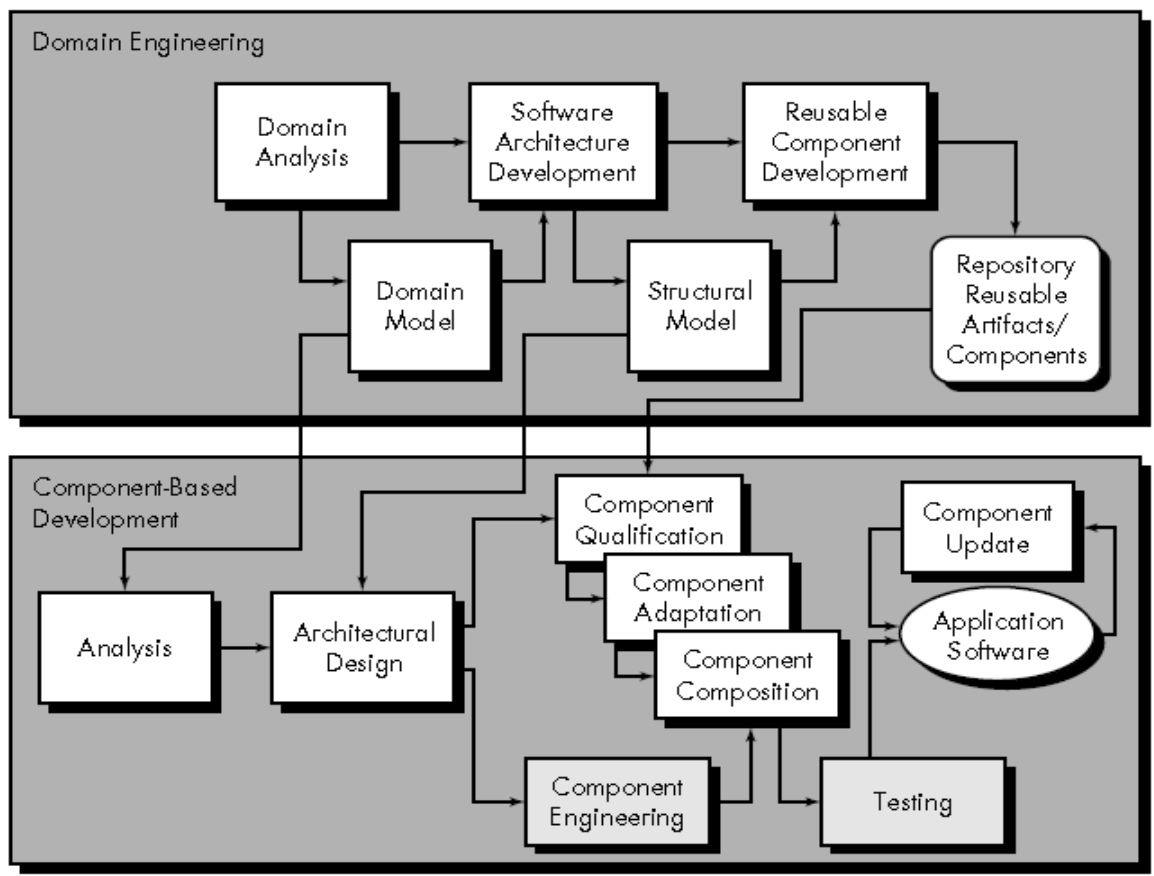

Figure: 1

The process model for component-based software engineering emphasizes parallel tracks in which domain engineering occurs concurrently with component-based development. *Domain engineering* performs the work required to establish a set of software components that can be reused by the software engineer. These components are then transported across a "boundary" that separates domain engineering from component-based development. Figure 1 illustrates a typical process model that explicitly accommodates CBSE. Domain engineering creates a model of the application domain that is used as a basis for analyzing user requirements in the software engineering flow. Generic software architecture provides input for the design of the application. Finally, after reusable components have been purchased, selected from existing libraries, or constructed (as part of domain engineering), they are made available to software engineers during component-based development.

9. **Domain Engineering:**

Domain engineering includes three major activities—analysis, construction, and dissemination. Domain construction and dissemination are considered in later sections in this chapter. It can be argued that "reuse will disappear, not by elimination, but by integration" into the fabric of software engineering practice . As greater emphasis is placed on reuse, some believe that domain engineering will become as important as software engineering over the next decade. According to Paul Clements Domain Engineering is about finding commonalities among system to identify components that can be applied to many systems and to identify program families that are positioned to take fullest advantage of those components.

**9.1 Domain Analysis Process (DAP):**

The steps in the process are:

1. Define the domain to be investigated
2. Categorize the items extracted from the domain
3. Collect a representative sample from the application
4. Analyze each application in the sample
5. Develop an analysis model for the objects

It is important to note that domain analysis is applicable to any software engineering paradigm and may be applied for conventional as well as object-oriented development. Prieto Diaz suggested an eight-step approach to the identification of the reusable components

1. Select specific functions or the objects
2. Abstract functions or object

3. Define a taxonomy
4. Identify common features
5. Identify specific relationship
6. Abstract the relationship
7. Derive a functional model
8. Define a domain language

Although the steps just noted provide a useful model for domain analysis, they provide no guidance for deciding which software components are candidates for reuse. Hutchinson and Hindley suggest the following set of pragmatic questions as a guide for identifying reusable software components:

• Is component functionality required on future implementations?
• How common is the component's function within the domain?
• Is there duplication of the component's function within the domain?
• Is the component hardware dependent?
• Does the hardware remain unchanged between implementations?
• Can the hardware specifics be removed to another component?
• Is the design optimized enough for the next implementation?
• Can we parameterize a nonreusable component so that it becomes reusable?
• Is the component reusable in many implementations with only minor changes?
• Is reuse through modification feasible?
• Can a nonreusable component be decomposed to yield reusable components?
• How valid is component decomposition for reuse?

**9.2 Characterization of Functions:**

It is sometimes difficult to determine whether a potentially reusable component is in fact applicable in a particular situation. To make this determination, it is necessary to define a set of domain characteristics that are shared by all software within a domain. A domain characteristic defines some generic attribute of all products that exist within the domain. For example, generic characteristics might include the importance of safety/reliability, programming language, concurrency in processing, and many others.

A set of domain characteristics for a reusable component can be represented by {Dp} where each item in the set has some specific domain characteristics namely $D_{pi}$

When new software, *w*, is to be built within the application domain, a set of domain characteristics is derived for it. A comparison is then made between *Dpi* and *Dwi* to determine whether the existing component *p* can be effectively reused in application *w*.

Even when software to be engineered clearly exists within an application domain, the reusable components within that domain must be analyzed to determine their applicability. In some cases (ideally, a limited number), "reinventing the wheel" may still be the most cost-effective choice.

### 9.3 Structure Modelling and Structure Points:

*Structural modeling* is a pattern-based domain engineering approach that works under the assumption that every application domain has repeating patterns (of function, data, and behaviour) that have reuse potential. According to Pollar and Rissman structural models consist of a small number of structural elements manifesting clear patterns of interaction. The architectures of systems using structural models are characterized by multiple ensembles that are composed from these model elements. Many architectural units emerge from simple patterns of interaction among this small number of elements". Structure Point is a distinct construct within a structure model.

McMahon [MCM95] describes a *structure point* as "a distinct construct within a structural model." Structure points have three distinct characteristics:

**1.** A structure point is an abstraction that should have a limited number of instances. Restating this in object-oriented jargon (Chapter 20), the size of the class hierarchy should be small. In addition, the abstraction should recur throughout applications in the domain. Otherwise, the cost to verify, document, and disseminate the structure point cannot be justified.

**2.** The rules that govern the use of the structure point should be easily understood. In addition, the interface to the structure point should be relatively simple.

**3.** The structure point should implement information hiding by isolating all complexity contained within the structure point itself. This reduces the perceived complexity of the overall system.

**Characterization of Structure points:**
1. It should have a limited number of instances
2. Interface of the structure points should be relatively simple
3. Structure point should implement information hiding by isolating all complexity contained within the structure point itself.

### 10. Component Engineering:

As we noted earlier in this chapter, the CBSE process encourages the use of existing software components. However, there are times when components must be engineered. That is, new software components must be developed and integrated with existing COTS and in-house components. Because these new components become members of the in-house library of reusable components, they should be engineered for reuse. Nothing is magical about creating software components that can be reused. Design concepts such as abstraction, hiding, functional independence, refinement, and structured programming, along with object-oriented methods, testing, SQA, and correctness verification methods, all contribute to the creation of software components that are reusable. Here we consider the reuse-specific issues that are complementary to solid software engineering practices.

## 11. Analysis and Design for Reuse:

Data, functional, and behavioral models (represented in a variety of different notations) can be created to describe what a particular application must accomplish. Written specifications are then used to describe these models. A complete description of requirements is the result. Ideally, the analysis model is analyzed to determine those elements of the model that point to existing reusable components. The problem is extracting information from the requirements model in a form that can lead to "specification matching." Bellinzoni, Gugini, and Pernici [BEL95] describe one approach for object-oriented systems: Components are defined and stored as specification, design, and implementation classes at various levels of abstraction—with each class being an engineered description of a product from previous applications. The specification knowledge—development knowledge—is stored in the form of reuse-suggestion classes, which contain directions for retrieving reusable components on the basis of their description and for composing and tailoring them after retrieval.

As we have already noted, DFR requires the software engineer to apply solid software design concepts and principles. But the characteristics of the application domain must also be considered. Binder [BIN93] suggests a number of key issues that form a basis for design for reuse:

**Standard data:** The application domain should be investigated and standard global data structures (e.g., file structures or a complete database) should be identified. All design components can then be characterized to make use of these standard data structures.

**Standard interface protocols:** Three levels of interface protocol should be established: the nature of intra-modular interfaces, the design of external technical (nonhuman) interfaces, and the human/machine interface.

**Program templates:** The structure model can serve as a template for the architectural design of a new program.

## 12. Software Lifecycle Process Model:

Lifecycle processes include all activities of a product or a system during its entire life, from the business idea for its development, through its usage and its completion of use. Different models have been proposed and exploit in software engineering, and different models have exhibit their (in)abilities to efficiently govern all activities required for a successful development and use of products. We can distinguish two main groups of models: Sequential and evolutionary. The sequential models define a sequence of activities in which one activity follow after a completion of the previous one. Evolutionary models allow performance of several activities in parallel without requirements on a stringent completion of one activity to be able to start with another one. Well known example of sequential models are waterfall model, or V model, and of evolutionary models, iterative and incremental development, or spiral model. CBSE addresses challenges similar to those encountered elsewhere in software engineering. Many of the methods, tools and principles of software engineering used in other types of system will be used in the same or a similar way in CBSE. There is however one difference; CBSE specifically focuses on questions related to components and in that sense it distinguishes the process of "component development" from that of "system development with components".

1. *Building System from Components*: The main idea of the component-based approach is building systems from pre-existing components. This assumption has several consequences for the system lifecycle. First, the development processes of component-based systems are separated from development processes of the components; the components should already been developed and possibly used in other products when the system development process starts. Second, a new separate process will appear: Finding and evaluating the components. Third, the activities in the processes will be different from the activities in non-component- based approach; for the system development the emphasis will be on finding the proper components and verifying them, and for the component development, design for reuse will be the main concern. There is a difference in requirements and business ideas in these two cases and different approaches are necessary. Components are built to be used and reused in many applications, some possibly not yet existing, in some possibly unforeseen way System development with components is focused on the identification of reusable entities and relations between them, beginning from the system requirements and from the availability of components already existing. We use V model as this model is widely used in many organizations – typically large organization building complex long-life products, such as cars or robots. In this model the process starts in a usual way by requirements engineering and requirements specification, followed by system specification. In a non- component- based approach the process would continue with the unit design, implementation and test. Instead of performing this activities that often are time and efforts consuming, we simply select appropriate components and integrate them in the system. However, two problems appear here which break this simplicity: (i) It is not obvious that there is any component to select, and (ii) the selected component only partially fits to our overall design. The first fact shows that we must have a process for finding components. This process includes activities for finding the components, and then the component evaluation. The second fact indicates for a need of component adoption and testing before it can be integrated into the system. And of course there must be a process of component development, this being independent of the system development process.

2. *Requirement Analysis and Specification*: In this phase one important activity is to analyse the possibility of realizing the solutions that will meet these requirements. In a component-based approach this implies that it is necessary to analyze whether these requirements can be fulfilled by available components. This means that the requirements engineers must be aware of components that can possibly be used. Since it is not likely that appropriate components can always be found, there is a risk that the new components have to be implemented. To keep with component-based approach (and utilize its advantages) one possibility is to negotiate the requirements and modify them to be able to use the existing components.

3. *System and Software Design:* Similar to the requirements specification phase the system specification and design is strongly related to the availability of the components. The potential components are complying with a particular component model. One could assume that it would be possible to used components implemented in different component technologies, but in practice it is very difficult to achieve interoperability between different component models. Particular component model requires a particular architectural framework, and the application is supposed to use this framework. This directly has impact on architectural decisions. For example if the component model requires a client-server architecture style, it is obvious that the application will use that style and not another (for example pipe-filter). This will put limitations on the system design. Also, other properties of components can have a direct influence on the design decisions. For this reason the design process is tightly connected to the availability of the components.

4. *Implementation and unit testing:* When building component-based system, an ideal case is to build an application by direct integration of components, i.e. directly connecting components. The "glue code" is a code that specifies this connection. In practice the role of the glue code will also include adaptation of the components, and even implementation of new functions. In an ideal case the components themselves are already built and tested. However the component tests in isolation are not sufficient. Often design units will be implemented as assemblies of several components and possibly a glue code. These assemblies must be tested separately, since an assembly of correct components may be incorrect although the components themselves are correct.

5. *System Integration*: The integration process includes integration of standard infrastructure components that build a component framework and the application components. The integration of a particular component into a system is called a component deployment. In difference to the entire system integration a component deployment is a mechanism for integration of particular components – it includes download and registering of the component.

6. *System Verification and validation:* The standard test and verification techniques are used here. The specific problem for component-based approach is location of error, especially when components are of "black box" type and delivered from different vendors. Typically a component can exhibit an error, but the cause of the malfunction lies in another component. Contractual interfaces play an important role in checking the proper input and output from components. These interfaces enable a specification of input and output and checking the correctness of data.

7. *Operation and Maintenance*: The maintenance process includes some steps that are similar to the integration process: A new or modified component is deployed into the system. Also it may be necessary to change the glue code. In most of the cases an existing component will be modified or a new version of the same component will be integrated into the system. Once again new problems caused by incompatibility between components, or by broken dependencies may occur. This means, one again that the system must be verified (either formally, or by simulation, or by testing).

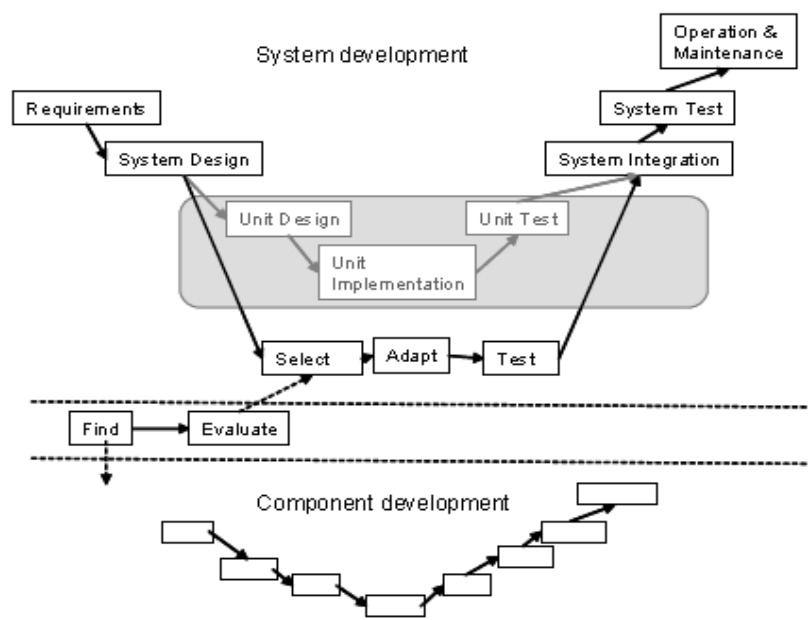

13. **Component Based Software Engineering Framework Activities**:
   If the requirement(s) cannot be changed or deleted, conventional or object-oriented software engineering methods are applied to develop those new components that must be engineered to meet the requirement(s). But for those requirements that are addressed with available components, a different set of software engineering activities commences:

   a) Component Qualification: System requirements and architecture define the components that will be required. Reusable components (whether COTS or in-house) are normally identified by the characteristics of their interfaces. That is, "the services that are provided, and the means by which consumers access these services" [BRO96] are described as part of the component interface. But the interface does not provide a complete picture of the degree to which the component will fit the architecture and requirements. The software engineer must use a process of discovery and analysis to qualify each component's fit.

   b) Component Adaptation: we noted that software architecture represents design patterns that are composed of components (units of functionality), connections, and coordination. In essence the architecture defines the design rules for all components, identifying modes of connection and coordination. In some cases, existing reusable components may be mismatched to the architecture's design rules. These components must be adapted to meet the needs of the architecture or discarded and replaced by other, more suitable components.

   c) Component Composition: Architectural style again plays a key role in the way in which software components are integrated to form a working system. By identifying connection and coordination mechanisms (e.g., run-time properties of the design), the architecture dictates the composition of the end product.

   d) Component Update: When systems are implemented with COTS components, update is complicated by the imposition of a third party (i.e., the organization that developed the reusable component may be outside the immediate control of the software engineering organization).

14. **Component Software Proposed by different Industries**: Because the potential impact of reuse and CBSE on the software industry is enormous, a number of major companies and industry consortia3 have proposed standards for component software:
   a) <u>OMG/CORBA</u>: The Object Management Group has published a *common object request broker architecture* (OMG/CORBA). An object request broker (ORB) provides a variety on services that enable reusable components (objects) to communicate with other components, regardless of their location within a system. When components are built using the OMG/CORBA standard, integration of those components (without modification) within a system is assured if an *interface definition language* (IDL) interface is created for every component. Using a client/server metaphor, objects within the client application request one or more services from the ORB server. Requests are made via an IDL or dynamically at run time. An interface repository contains all necessary information about the service's request and response formats.

b) <u>Microsoft COM:</u> Microsoft has developed a component object model (COM) that provides a specification for using components produced by various vendors within a single application running under the Windows operating system. COM encompasses two elements: COM interfaces (implemented as COM objects) and a set of mechanisms for registering and passing messages between COM interfaces. From the point of view of the application, "the focus is not on how [COM objects are] implemented, only on the fact that the object has an interface that it registers with the system, and that it uses the component system to communicate with other COM objects."

c) <u>SUN JavaBeans Components:</u> The JavaBean component system is a portable, platform independent CBSE infrastructure developed using the Java programming language. The JavaBean system extends the Java applet4 to accommodate the more sophisticated software components required for component-based development. The JavaBean component system encompasses a set of tools, called the *Bean Development Kit* (BDK) that allows developers to (1) analyse how existing Beans (components) work, (2) customize their behaviour and appearance, (3) establish mechanisms for coordination and communication. (4) Develop custom Beans for use in a specific application, and (5) test and evaluate Bean behaviour.

15. **Common Object Request Broker Architecture**:

    The Common Object Request Broker Architecture (CORBA) is a standard defined by the Object Management Group (OMG) that enables software components written in multiple computer languages and running on multiple computers to work together (i.e., it supports multiple platforms).

    CORBA is useful because it enables separate pieces of software written in different languages and running on different computers to work together like a single application or set of services. More specifically, CORBA is a mechanism in software for normalizing the method-call semantics between application objects residing either in the same address space (application) or remote address space (same host, or remote host on a network). Version 1.0 was released in October 1991. CORBA uses an interface definition language (IDL) to specify the interfaces which objects present to the outer world. CORBA then specifies a *mapping* from IDL to a specific

implementation language like C++ or Java. Standard mappings exist for Ada, C, C++, Lisp, Ruby, Smalltalk, Java, COBOL, PL/I and Python. There are also non-standard mappings for Perl, Visual Basic, Erlang, and Tcl implemented by object request brokers (ORBs) written for those languages.

The CORBA specification dictates there shall be an ORB through which an application would interact with other objects. In practice, the application simply initializes the ORB, and accesses an internal *Object Adapter*, which maintains things like reference counting, object (and reference) instantiation policies, and object lifetime policies. The Object Adapter is used to register instances of the *generated code classes*. Generated code classes are the result of compiling the user IDL code, which translates the high-level interface definition into an OS- and language-specific class base for use by the user application. This step is necessary in order to enforce CORBA semantics and provide a clean user process for interfacing with the CORBA infrastructure.

Some IDL language mappings are "more hostile" than others. For example, due to the nature of Java, the IDL-Java Mapping is rather straightforward and makes usage of CORBA very simple in a Java application. The C++ mapping is less straightforward, but it accounts for all CORBA features (e.g., exception handling). The C mapping is even stranger (since C is not an object-oriented language), but it does make sense and properly handles the RPC semantics.

A language mapping requires the developer ("user" in this case) to create some IDL code that represents the interfaces to his objects. Typically, a CORBA implementation comes with a tool called an IDL compiler which converts the user's IDL code into some language-specific generated code. A traditional compiler then compiles the generated code to create the linkable-object files for the application. This diagram illustrates how the generated code is used within the CORBA infrastructure:

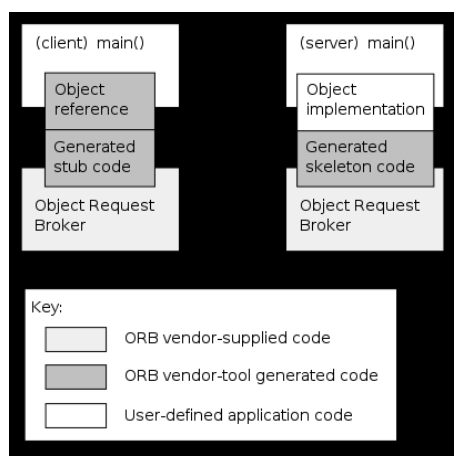

This figure illustrates the high-level paradigm for remote interposes communications using CORBA. Issues not addressed here, yet accounted-for in the CORBA specification include: data typing, exceptions, network protocol, communication timeouts, etc. For example: Normally the server side has the Portable Object Adapter (POA) that redirects call either to the local servants or (to balance the load) to the other servers. Also, both server and client parts often have interceptors that are described below. Issues CORBA (and thus this figure) do not address, but that all distributed systems must address: object lifetimes, redundancy/fail-over, naming semantics (beyond a simple name), memory management, dynamic load balancing, separation of model between display/data/control semantics, etc.

*Limitations of Standard CORBA Model:*
- *No standard way to deploy object implementations*: It did not define a standard for deployment of object implementations in server processes.

- *Limited standard support for common CORBA server programming patterns*: Provides a rich set of features to implement servers
- Limited extension of object functionality: In the traditional CORBA object model, objects can be extended only via inheritance and multiple inheritance cannot expose the same interface more than once, nor can it alone determine which interface should be exported to clients
- Availability of CORBA object Server is not defined in advance
- No standard Object Life cycle Management

### 16. CORBA Component Model (CCM):

CORBA Component Model (CCM) is an addition to the family of CORBA definitions. It was introduced with CORBA 3 and it describes a standard application framework for CORBA components. Though not dependent on "language independent Enterprise Java Beans (EJB)", it is a more general form of EJB, providing four component types instead of the two that EJB defines. It provides an abstraction of entities that can provide and accept services through well-defined named interfaces called *ports*.

The CCM has a component container, where software components can be deployed. The container offers a set of services that the components can use. These services include (but are not limited to) notification, authentication, persistence and transaction processing. These are the most-used services any distributed system requires, and, by moving the implementation of these services from the software components to the component container, the complexity of the components is dramatically reduced.

*Disadvantages of CCM Model:*

❏ CCM Specifications are very large and complex
❏ As with CORBA specifications several years ago, it is still hard to evaluate the quality and performance of CCM implementations
❏ Interlinking of Components and the container are very difficult to understood

17. **Classifying and Retrieving Components:** Consider a large university library. Tens of thousands of books, periodicals, and other information resources are available for use. But to access these resources, a categorization scheme must be developed. To navigate this large volume of information, librarians have defined a classification scheme that includes a Library of Congress classification code, keywords, author names, and other index entries. All enable the user to find the needed resource quickly and easily. Now, consider a large component repository. Tens of thousands of reusable software components reside in it. But how does a software engineer find the one she needs? To answer this question, another question arises: How do we describe software components in unambiguous, classifiable terms? These are difficult questions, and no definitive answer has yet been developed. In this section we explore current directions that will enable future software engineers to navigate reuse libraries.

Describing Reusable Component: A reusable software component can be described in many ways, but an ideal description encompasses what Tracz [TRA90] has called the *3C model*—concept, content and context. The *concept* of a software component is "a description of what the component does" [WHI95]. The interface to the component is fully described and the semantics— represented within the context of pre- and postconditions—are identified. The concept should communicate the intent of the component. The *content* of a component describes how the concept is realized. In essence, the content is information that is hidden from casual users and need be known only to those who intend to modify or test the component. The *context* places a reusable software component within its domain of applicability. That is, by specifying conceptual, opera tional, and implementation features, the context enables a software engineer to find the appropriate component to meet application requirements.

*Classification Scheme:*

Different papers have been published based on classification scheme. The majority of classification schemes for software components fall into three categories:

a) *Enumerated Classification*: Components are described by a hierarchical structure in which classes and varying levels of subclasses of software components are defined. Actual components are listed at the lowest level of any path in the enumerated hierarchy. The hierarchical structure of an enumerated classification scheme makes it easy to understand and to use. However, before a hierarchy can be built, domain engineering must be conducted so that sufficient knowledge of the proper entries in the hierarchy is available.

b) *Faceted Classification*:  A domain area is analyzed and a set of basic descriptive features are identified. These features, called *facets,* are then ranked by importance and connected to a component. A facet can describe the function that the component performs, the data that are manipulated, the context in which they are applied, or any other feature. The set of facets that describe a component is called the *facet descriptor.* Generally, the facet description is limited to no more than seven or eight facets. As a simple illustration of the use of facets in component classification, consider a scheme [LIA93] that makes use of the following facet descriptor: {function, object type, system type} Each facet in the facet descriptor takes on one or more values that are generally descriptive keywords. For example, if *function* is a facet of a component, typical values assigned to this facet might be *function* = (*copy, from*) or (*copy, replace, all*)

c) *Attribute-value Classification*: A set of attributes is defined for all components in a domain area. Values are then assigned to these attributes in much the same way as faceted classification. In fact, attribute value classification is similar to faceted classification with the following exceptions: (1) no limit is placed on the number of attributes that can be used; (2) attributes are not assigned priorities, and (3) the thesaurus function is not used.

18. **Reuse Environment:**  Software component reuse must be supported by an environment that encompasses the following elements:
    • A component database capable of storing software components and the classification information necessary to retrieve them.
    • A library management system that provides access to the database.
    • A software component retrieval system (e.g., an object request broker) that enables a client application to retrieve components and services from the library server.

• CBSE tools that support the integration of reused components into a new design or implementation.

A component classification scheme serves as the basis for library queries. Queries are often characterized using the context element of the 3C model described earlier in this section. If an initial query results in a voluminous list of candidate components, the query is refined to narrow the list. Concept and content information are then extracted (after candidate components are found) to assist the developer in selecting the proper component.

19. **Impact on Quality, Productivity and Cost:** Component-based software engineering has an intuitive appeal. In theory, it should provide a software organization with advantages in quality and timeliness. And these should translate into cost savings. But are there hard data that support our intuition? To answer this question we must first understand what actually can be reused in a software engineering context and then what the costs associated with reuse really are. As a consequence, it is possible to develop a cost/benefit analysis for component reuse.

    a) *Quality:*  In an ideal setting, a software component that is developed for reuse would be verified to be correct and would contain no defects. In reality, formal verification is not carried out routinely, and defects can and do occur. However, with each reuse, defects are found and eliminated, and a component's quality improves as a result. Over time, the component becomes virtually defect free. In a study conducted at Hewlett Packard, Lim [LIM94] reports that the defect rate for reused code is 0.9 defects per KLOC, while the rate for newly developed software is 4.1 defects per KLOC. For an application that was composed of 68 percent reused code, the defect rate was 2.0 defects per KLOC—a 51 percent improvement from the expected rate, had the application been developed without reuse. Henry and Faller [HEN95] report a 35 percent improvement in quality. Although anecdotal reports span a reasonably wide spectrum of quality improvement percentages, it is fair to state that reuse provides a nontrivial benefit in terms of the quality and reliability for delivered software.
    b) *Productivity:*  When reusable components are applied throughout the software process, less time is spent creating the plans, models, documents, code, and data that are required to create a deliverable system. It follows that the same level of functionality is delivered to the customer with less input effort. Hence, productivity is improved. Although percentage productivity improvement reports are notoriously difficult to interpret,8 it appears that 30 to 50 percent reuse can result in productivity improvements in the 25–40 percent range.
    c) *Cost:*  The net cost savings for reuse are estimated by projecting the cost of the project if it were developed from scratch, $C_s$, and then subtracting the sum of the

costs associated with reuse, *Cr*, and the actual cost of the software as delivered, *Cd*. *Cs* can be determined by applying one or more of the estimation techniques. The costs associated with reuse, *Cr*, include
- Domain analysis and modeling.
- Domain architecture development.
- Increased documentation to facilitate reuse.
- Support and enhancement of reuse components.
- Royalties and licenses for externally acquired components.
- Creation or acquisition and operation of a reuse repository.
- Training of personnel in design and construction for reuse.

Although costs associated with domain analysis and the operation of a reuse repository can be substantial, many of the other costs noted here address issues that are part of good software engineering practice, whether or not reuse is a priority.

20. **Cost Analysis Using Structure Points:** we defined a structure point as an architectural pattern that recurs throughout a particular application domain. A software designer (or system engineer) can develop an architecture for a new application, system, or product by defining a domain architecture and then populating it with structure points. These structure points are either individual reusable components or packages of reusable components. Since all structure points (and reusable components in general) have a past history, cost data can be collected for each. In an ideal setting, the qualification, adaptation, integration and maintenance costs associated with each component in a reuse library is maintained for each instance of usage. These data can then be analyzed to develop projected costs for the next instance of reuse.

- Consider a new application, *X*, that requires 60 percent new code and the reuse of three structure points, SP1, SP2, and SP3. Each of these reusable components has been used in a number of other applications and average costs for qualification, adaptation, integration, and maintenance are available.
- To estimate the effort required to deliver *X*, the following must be determined:
    overall effort = $E$new + $E$qual + $E$adapt + $E$int
  where
  $E$new = effort required to engineer and construct new software components
  $E$qual = effort required to qualify SP1, SP2, and SP3.
  $E$adapt = effort required to adapt SP1, SP2, and SP3.
  $E$int = effort required to integrate SP1, SP2, and SP3.
  The effort required to qualify, adapt, and integrate SP1, SP2, and SP3 is determined by taking the average of historical data

**21. Disadvantages of CBD:**

The main disadvantages of CBD are given below:

1. it is very difficult to build the environment that is fitted to component
2. Concept of reuse: Ideally speaking, standards are needed regarding middleware in which the component is supposed to work. (Middleware: A communication layer which enables components to interact with higher level component in a network). But among the three standards available, only CORBA is language independent. Hence the comparison is not viable among the standards.

**22. Advantages of CBD:**

1. flexibility: runtime components can work independently if properly designed and they are less dependent on the environment
2. Reuse: Once developed, it can be used everywhere regarding the programming language and OS. But domain engineering should be kept in mind
3. Easy to maintain because ideally the functionality is implemented once
4. development cost is much lower
5. Lesser time required to build the software

**23. Advancement in Component Based Approach:**

Component-based Software Engineering (CBSE) has emerged as a technology for the rapid assembly of flexible software systems. CBSE combines elements of software architecture, modular software design, software verification, configuration and deployment. To foster exchange and collaboration with the software architecture community, CBSE is colocated with the Quality of Software Architectures Conference (QoSA) and the International Symposium on Architecting Critical Systems (ISARCS)as part of the federated CompArch event.

The theoretical foundations of component specification, composition, analysis and verification continue to pose research challenges. While the engineering models and methods for component software development are slowly maturing, new trends in global services, distributed systems architectures, and large scale software systems that cross organizational boundaries push the limits of established and tested component-based methods, tools and platforms:

- model-driven development and grid technologies with their high-performance demands in massive data storage, computational complexity and global co-scheduling of scientific models in flagship science, technology and medicine research;

- global software development with its lowering of cost of software capabilities and production, through automation, off-shoring and outsourcing of key components and subsystems;
- networked enterprise information systems and services architectures crossing enterprise, nation, legal and discipline boundaries;
- shift from (globally distributed) software products to pervasive and ubiquitous services supported by deep software-intensive infrastructures and middleware and by increasingly flexible, adaptive and autonomous client and application server software.

In the CBSE 2010 technical symposium following topics were discussed based on recent advancement in component based approach.

- Design of component models
- Theories (including taxonomies) of software composition and binding
- Coordination and choreography of component software, services, workflows
- Run-time adaptation of component-based systems
- Interaction between component models, software architectures and product lines
- Component-based web services and service-oriented architecture
- Declarative, rule-based management of component-based systems
- Software quality and extra-functional properties for components and component-based systems
- Global generation, adaptation and deployment of component-based systems and services
- Components and generative approaches
- Components and model-driven development
- Specification, verification and testing of component-based systems
- Compositional reasoning techniques for component models
- Global measurement, prediction and monitoring of distributed and service components
- Patterns and frameworks for component-based systems and services
- Integrated tool chains and methods for building component-based services
- Components for networked real-time information systems and sensor networks;
- Industrial experience using component-based software development;
- Empirical studies in component-based software engineering;
- Teaching component-based software engineering

## 24. Organization doing CBD:

Relatively few organizations have started doing CBD seriously. It is being adopted faster in some industries than others - notably insurance and other financial sectors, and telecoms. Many other organizations are holding a watching brief - they are attending industry and vendor briefings, or joining the CBD Forum, but have not yet started doing CBD seriously. A number of large software vendors have made a major commitment to Component-Based Development, including Forte, IBM, Microsoft, SAP, Sterling and Sun

## 25. Summary:

Component-based software engineering offers inherent benefits in software quality, developer productivity, and overall system cost. And yet, many roadblocks remain to be overcome before the CBSE process model is widely used throughout the industry. In addition to software components, a variety of reusable artifacts can be acquired by a software engineer. These include technical representations of the software (e.g., specifications, architectural models, designs), documents, test data, and even processrelated tasks (e.g., inspection techniques). The CBSE process encompasses two concurrent subprocesses—domain engineering and component-based development. The intent of domain engineering is to identify, construct, catalog, and disseminate a set of software components in a particular application domain. Component-based development then qualifies, adapts, and integrates these components for use in a new system. In addition component-based development engineers new components that are based on the custom requirements of a new system, Analysis and design techniques for reusable components draw on the same principles and concepts that are part of good software engineering practice. Reusable components should be designed within an environment that establishes standard data structures, interface protocols, and program architectures for each application domain.

Few key points regarding CBSE is mentioned below:
1. CBSE Process has two sub-process – Domain engineering and CBD.
2. CBD has 4 activities – Qualification, Adaptation, composition, updation.
3. V-model is used for the component based software lifecycle.
4. The object model generally conforms to one or more standards (OMG/CORBA, COM, DCOM, JavaBean).
5. Classification schemes enable the developer to find and retrieve the reusable components
6. CORBA object model is increasingly gaining acceptance as the industry standard, cross-platform, cross-language distributed object computing model
7. The CCM programming model is suitable for existing service to develop the next generation of highly scalable distributed application

8. A component based approach can not be utilized if the development processes are not adopted according to CBSE principles
9. CBD has a lot of promises but is not silver bullet